# Early Diagnosis of Parkinson's Disease by Analyzing Magnetic Resonance Imaging Brain Scans and Patient Characteristics

**Sabrina Zhu**
**The Harker School, 500 Saratoga Ave, San Jose, 95129 CA**

**Abstract**

Parkinson's disease (PD) is a chronic condition that affects motor skills and includes symptoms like tremors and rigidity. The current diagnostic procedure uses patient assessments to evaluate symptoms and sometimes a magnetic resonance imaging (MRI) scan. However, symptom variations cause inaccurate assessments, and the analysis of MRI scans requires experienced specialists. This research proposes to accurately diagnose PD severity with deep learning by combining symptoms data and MRI data from the Parkinson's Progression Markers Initiative (PPMI) database. A new hybrid model architecture was implemented to fully utilize both forms of clinical data, and models based on only symptoms and only MRI scans were also developed. The symptoms-based model integrates a fully connected deep learning neural network, and the MRI scans- and hybrid models integrate transfer learning-based convolutional neural networks. Instead of performing only binary classification, all models diagnose patients into five severity categories, with stage zero representing healthy patients and stages four and five representing patients with PD. The symptoms-only, MRI scans-only, and hybrid models achieved accuracies of 0.77, 0.68, and 0.94, respectively. The hybrid model also had high precision and recall scores of 0.94 and 0.95. Real clinical cases confirm the hybrid model's strong performance, where patients were classified incorrectly with both other models but correctly by the hybrid. It is also consistent across the five 0-4 severity stages, indicating accurate early detection. This is the first report to combine symptoms data and MRI scans with a machine learning approach on such a large scale.

# 1. Introduction

Parkinson's disease is a chronic and progressive condition that mainly affects a patient's motor skills and can include symptoms such as tremors, rigidity, slowness of movement or bradykinesia, and difficulty with walking and coordination (National Institute of Health's National Institute on Aging). The disease can also present itself with changes in mental behavior, including sleep problems, fatigue, depression, and anxiety. Approximately 1.04 million individuals in the United States have Parkinson's disease, based on a 2017 population (Yang et al. 2020). In 2010, there were about 680,000 patients in the US, and in 2030, this number is predicted to rise to 1,238,200 (Marras et al. 2018).

Currently, in clinical practices, doctors diagnose Parkinson's patients by looking for certain characteristics such as movement-limiting symptoms (American Association of Neurological Surgeons) with checklists like the Movement Disorder Society (MDS) Clinical Diagnostic Criteria for Parkinson's disease (Postuma et al. 2015). Sometimes, medical history and neurological examinations follow (American Parkinson Disease Association). However, the disease can appear differently for different patients, so only analyzing symptoms with behavioral assessments can be inaccurate. Doctors can also require magnetic resonance imaging (MRI) scans to be done as a part of the clinical diagnosis process. MRI scans use noninvasive radio waves and a magnetic field to analyze body tissues, and they can provide detailed and accurate visualizations of human bodies. MRI scans of human brains provide another perspective in distinguishing between different types of parkinsonism, including PD and various atypical parkinsonism disorders (Heim et al. 2017). Diagnosis of PD is still inaccurate; over about the last three decades, diagnostic accuracy is estimated to be only 80.6% (Rizzo et al. 2016).

With recent advancements in machine learning, models have greatly aided in the diagnosis of brain conditions such as Parkinson's disease. A summary of previous work in this field is shown in Table 1. Recently, researchers have also begun to apply deep learning to diagnose patients with Parkinson's disease. Machine learning models have been relatively successful (Saha 2019) (Sivaranjini et al. 2019) (Haller et al. 2012). These studies analyze MRI scans with models like the convolutional neural network (CNN) and the support vector machine (SVM) model. Three-dimensional (3D) CNNs have also been explored (Chakraborty et al. 2020) (Esmaeilzadeh et al. 2018). Furthermore, some studies have used only patient scores or symptom-based assessments (Prashanth et al. 2016), along with other biomarkers. These studies use a variety of machine learning models and data types, both of which are factors explored in this research.

| Paper | Models | Results |
|---|---|---|
| Prashanth et al. 2016 | Developed a naive bayes, support vector machine, and boosted trees and random forests algorithms, classified subjects into two categories | Reached an accuracy of 96.40% |
| Saha et al. 2019 | Developed a convolutional neural network, classified subjects into two categories | Reached an accuracy of 97.91% |
| Archer et al. 2019 | Developed support vector machine algorithms with MRI and the MDS-UPDRS III test, MRI only, and MDS-UPDRS III only, classified subjects into categories of Parkinsonism | Reached accuracies of 91.49% with MRI and MDS-UPDRS III model; 90.24% with MRI only model; 66.61% with MDS-UPDRS III only model |
| Shu et al. 2020 | Developed a maximum relevance minimum redundancy algorithm, analyzed patients in Parkinson's disease stages 1 and 2 | Reached accuracies of 82.7% for stage 1 and 85.4% for stage 2 |

Table 1: Summary of previous work in diagnosis of Parkinson's disease with MRI scans and machine learning.

While MRI scans can provide useful information about a patient's condition, they can be expensive and are unavailable in some underdeveloped areas. Additionally, the symptom-based data assessments measure is also incredibly important (National Institute of Health's National Institute on Aging). Unlike cancer or bone injuries, which can be diagnosed by experienced doctors' studying scans, it is not as easy to rely only on MRI scans for PD diagnosis and classification because there are no obvious visible discrepancies (Pagano et al. 2016). However, to the best of my knowledge, only two studies so far (Archer et al. 2019) (Abel et al. 2019) have incorporated patient scores into their analyses of MRI scans, and they each only used one patient assessment score. The assessment both studies used, the MDS Unified Parkinson's Disease Rating Scale Part III, measures motor skills and is one of the most commonly used and accepted assessments in the diagnosis of Parkinson's disease (Postuma et al. 2015). In contrast, our study incorporates these symptoms along with many others, including non-motor dysfunction and rapid eye movement (REM) disorders, since PD patients often experience a much wider variety of symptoms. Furthermore, most previous studies have only performed binary classification, separating patients into either having the disease or not having the disease (Haller et al. 2012) (Prashanth et al. 2016) (Archer et al. 2019). However, binary classification is limited in usefulness because Parkinson's disease can appear very differently across different patients. Therefore, it is also important to determine the severity of the condition, which we address by performing multi-category classification (National Institute of Health's National Institute on Aging). Different medication and treatment may be required for patients with milder or more severe symptoms, so it is critical that the Parkinson's disease stage is calculated accurately.

We aim to explore the possibilities of using deep learning and convolutional neural networks to diagnose Parkinson's disease with a combination of assessment score data and MRI scans data.

We hypothesize that there would be multiple benefits to our models since we incorporate different and more types of data and also classify patients into five categories, based on the severity of the disease, instead of just two. Both forms of data have been incorporated, and a model that is a close digital representation of the diagnostic process of human doctors analyzing Parkinson's disease patients was designed. A model based solely on symptom assessments and a model based solely on MRI scans were also developed, as shown in Figure 1. The goal of this research was to build customized neural networks that can provide diagnoses with highly accurate performance. This is the first report to combine both forms of data in a machine learning based diagnosis approach with such a large scale amount of symptom assessment scores.

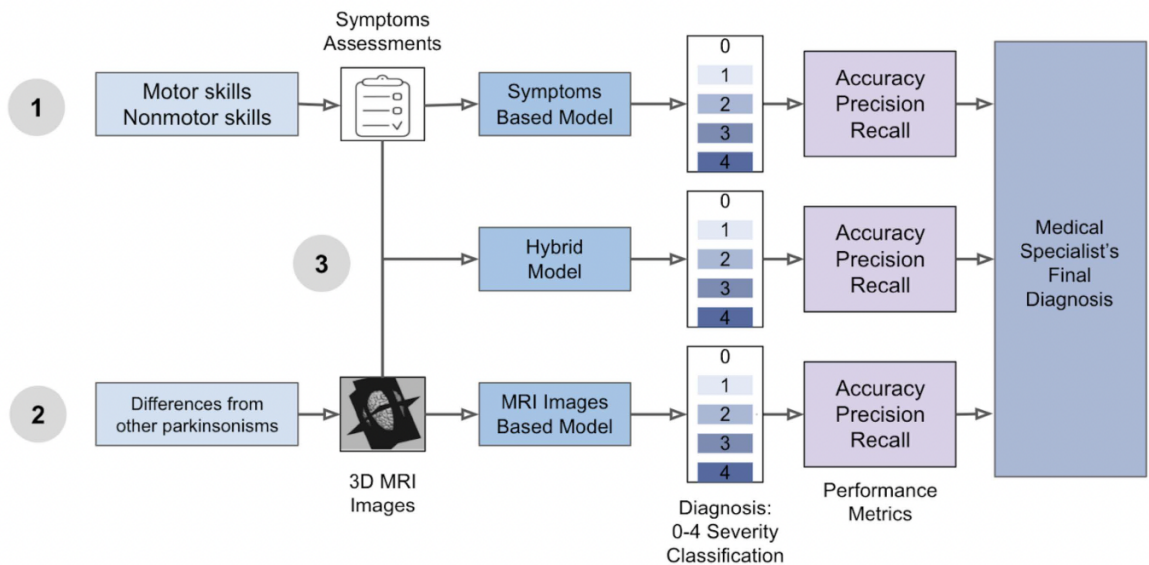

Figure 1: Proposed diagnosis method for Parkinson's pre-screening based on newly developed hybrid model which fully utilizes two forms of data: numerical symptoms assessments and MRI images.

## 2. Proposed Methods

### 2.1 Classification of Parkinson's Disease Patients

Parkinson's disease patients are classified into six different categories based on the severity of the disease. The Hoehn and Yahr Scale Assessment (HYR), developed in 1967, measures the level of disability and the progression of Parkinson's disease and is useful in current diagnostic processes (Hoehn et al. 1967) (Goetz et al. 2004). Stage zero represents a healthy patient, stage three represents a patient with early and mild signs of Parkinson's disease, and stage five represents a symptomatic patient diagnosed with Parkinson's disease. In this research, stages four and five were combined, since there were only a few patients in each. Most individuals in the dataset used in this research have already been assigned a value from the HYR. Table 2 gives descriptions of each stage.

| Stage | Symptoms |
|---|---|
| 0 | Healthy; no symptoms |
| I | Unilateral symptoms; minor impairment |
| II | Bilateral symptoms; no balance impairment |
| III | Difficulty with balance; moderate disability |
| IV&V | Severe symptoms; confined to wheelchair |

Table 2: Classification stages of patients based on Hoehn & Yahr scale.

**2.2 Database for Deep Learning Training**

The Parkinson's Progression Markers Initiative (PPMI) is an observational clinical study used to verify progression markers in Parkinson's disease and is being conducted at a network of clinical sites in the United States, Europe, Australia, and Israel. It establishes a comprehensive set of clinical, imaging, and biosample data used to define biomarkers of PD progression, and the database has been used by multiple groups to study the disease (Prasanth et al. 2016) (Saha 2019) (Chakraborty et al. 2020). The dataset contains 2952 MRI scans of PD patients and 925 MRI scans of healthy controls, who have no reported history of medical conditions. Scans were taken with a Siemens Trio 3 Tesla MRI scanner and 3D magnetization-prepared rapid acquisition with gradient echo (MPRAGE).

**2.3 Data Preprocessing**

Besides the HYR measurements, the PPMI dataset also includes forty other tests, which assess qualities like patient motor skills and non-motor skills. Each of these tests contains multiple questions, usually yes or no, or features. To determine which features to use, we found the correlations between them. We chose 0.5 as the limit to ensure they were all relatively independent (Akoglu 2018). If a pair had a correlation greater than 0.5 or less than -0.5, we removed one of the features, and this resulted in 94 features. These 94 features came from a total of 10 tests: the Epworth Sleepiness Scale, which assesses the 'daytime sleepiness' of the patients, a symptom found in 50-75% of patients; the MDS UPDRS Part I, which analyzes the nonmotor aspects of experiences of daily living (Goetz et al. 2008); MDS UPDRS Part II, which measures the motor aspects of experiences of daily living; MDS UPDRS Part III, which is a motor examination; MDS UPDRS Part IV, which assesses motor complications; the Montreal Cognitive

Assessment (MoCA), which analyzes short term memory, visuospatial abilities, attention, and orientation (Nasreddine et al. 2005); the Olfactory University of Pennsylvania Smell Identification Test (UPSIT), which diagnoses olfactory disorder (Doty et al. 1984); the Physical Activity Scale for the Elderly (PASE) Household Activity, which measures the physical abilities of older adults (Washburn et al. 1993); the Rapid Eye Movement (REM) Sleep Disorder Questionnaire, which detects REM disorders (Stiasny-Kolster et al. 2007); and the Scales for Outcomes in Parkinson's Disease - Autonomic Dysfunction (SCOPA-AUT), which analyzes autonomic symptoms (Visser et al. 2004). Examples of the questions asked in these assessments include "Over the past week, have you had uncomfortable feelings in your body like pain, aches, tingling, or cramps?," whose answer options were 0: Normal, 1: Slight, 2: Mild, 3: Moderate, or 4: Severe and which was found in the MDS UPDRS test, and "My sleep is frequently disturbed," whose answer options were yes or no and which was found in the REM Sleep Disorder Questionnaire.

In addition to preprocessing the features, we also preprocessed the MRI scans. Classifying patient types from MRI scans can be viewed as an image classification problem, and CNNs are the dominant deep learning frameworks for image classification (O'Shea et al. 2015). However, "traditional" CNN's usually classify based on individual 2D images. Since the input data were 3D images, they were first converted to 2D images. Using BrainSuite, a software that helps visualize and analyze the brain (BrainSuite), the MRI scans were run through extraction stages, which removed the skull and corrected for nonuniformity and topology. The central point of the brain was located and cross-sections along the x, y, and z axes were taken. This resulted in a sagittal, coronal, and transverse image for each scan, which were then fitted into three "channels" to fit into the CNN framework. A sample scan and its corresponding images are displayed in Figure 2.

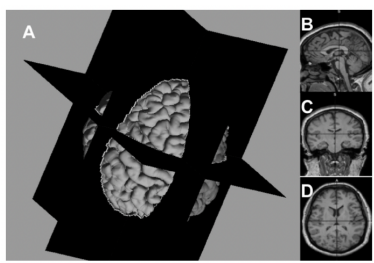

Figure 2: Sample 3D MRI scan of a Parkinson's disease patient with its three 2D image cross sections. (A) sample 3D MRI scan of Parkinson's disease patient; (B) 2D image cross section taken along sagittal axis; (C) 2D image cross section taken along coronal axis; (D) 2D image cross section taken along transverse axis.

| Stage | Number of Patients |
|-------|--------------------|
| 0     | 5                  |
| I     | 31                 |
| II    | 146                |
| III   | 12                 |
| IV&V  | 2                  |

Table 3: Number of patients in each classification category.

With the data, a symptoms-only based model was developed as well as an MRI scans-only based model. Finally, the hybrid model was developed, with both patient assessment scores and MRI scans as input data. To keep the data consistent, only patients who had all the assessment scores, as well as MRI scans, were selected. This resulted in 196 patients, each with full sets of assessment feature data and scans. Table 3 shows a breakdown of how many patients were in each category.

**2.4 Deep Learning Models**

**2.4.1 Symptoms Based Model**

In this research, symptoms assessment data was used to diagnose Parkinson's disease. First, a logistic regression model using the assessment scores was developed. Multinomial logistic regression was chosen as opposed to linear regression because patients were classified into categories. Furthermore, logistic regression is very popular and is commonly used for classification problems (Christodoulou et al. 2019). The model incorporated the 94 features and implemented two layers of weighted nodes. The patients were finally classified into five categories. 80% of the available data, or 397 sets of data, was used to train the model, and 20%, or 99 sets of data, for testing and validation.

**2.4.2 MRI Scans Based Model**

A second model based only on MRI scans was also designed to perform diagnoses for Parkinson's disease patients. A convolutional neural network was selected, as CNN's are optimal for image classification tasks. Instead of building a complete CNN model from scratch, transfer learning with the Google MobileNet V2 model (Sandler et al. 2018) was used. The MobileNet model has been trained on Imagenet data and is very accurate at detecting image features, such as edges and shapes. With transfer learning, the model is greatly simplified, and the training time is reduced. After transfer learning, two more convolution and pooling layers were added. The patients were finally classified into five categories. Training and testing data was divided in the same way as the symptoms-based model.

### 2.4.3 Hybrid Model

Finally, a model that fully takes advantage of all of the patient's available data, both symptom assessments and MRI scans, was developed. The hybrid model algorithm is structured mainly on the MRI scans based model, with assessment features being fed in during a convolutional layer. After running the 2D scans through the MobileNet V2 architecture and two layers of weighted nodes, the assessment scores were normalized and fed in. Then, the model trained on another convolutional layer before finalizing the classification output. Deep learning was then implemented with the hybrid model, and an additional layer of weighted notes for the assessment scores was inserted before integrating them with the MRI scans, as shown in Figure 3. The patients were finally classified into five categories. Training and testing data was divided in the same way as the symptoms-based model and MRI scans-based model.

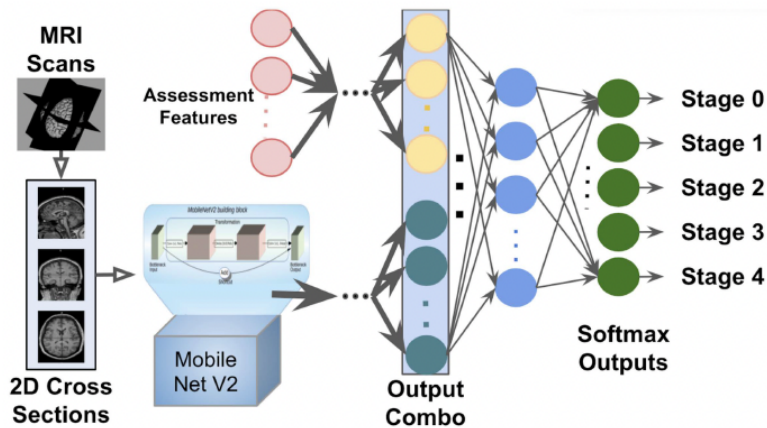

Figure 3: Diagram of design architecture of the proposed hybrid model. The model integrates the symptoms based and MRI images based models by merging their outputs into a single layer. With the complementing outputs, the hybrid model gives a 0-4 stage output as the final severity prediction.

## 3. Results and Discussion

### 3.1 Results from Symptoms Based Model

In this research, performance metrics of accuracy; precision, or the proportion of true positives in all positives classified by the model; and recall, or the proportion of true positives in all actual positives, were used to evaluate the models' performances. Initially, the symptoms-based model reached a training accuracy of 0.98, testing accuracy of 0.68, precision of 0.61, and recall of 0.69. In order to optimize the initial model, deep learning was applied; an additional layer was added before giving an output. This modification allowed the model to be trained better, and afterward, it reached a training accuracy of 0.98, testing accuracy of 0.77, precision of 0.68, and recall of 0.77, as shown in Table 4. The model performed better than previous models, including that of Archer et al 2019, where only one assessment was used and where only an accuracy of 0.66 was achieved. Because the 94 features come from a total of only ten tests, it will not be difficult for patients to complete the assessments and to retrieve sufficient data.

|  | Symptoms Based | MRI Scans Based | Hybrid |
| --- | --- | --- | --- |
| Model | Logistic Regression | CNN | MobileNet V2 CNN |
| Input Data | 92 symptoms | 2D MRI scan | 2D MRI scan, 92 symptoms |
| Number of Patients | 496 | 496 | 496 |
| Training Accuracy | 0.98 | 0.98 | 0.99 |
| Testing Accuracy | 0.77 | 0.68 | 0.94 |
| Precision | 0.68 | 0.69 | 0.94 |
| Recall | 0.77 | 0.74 | 0.95 |

Table 4: Summary of results from different models.

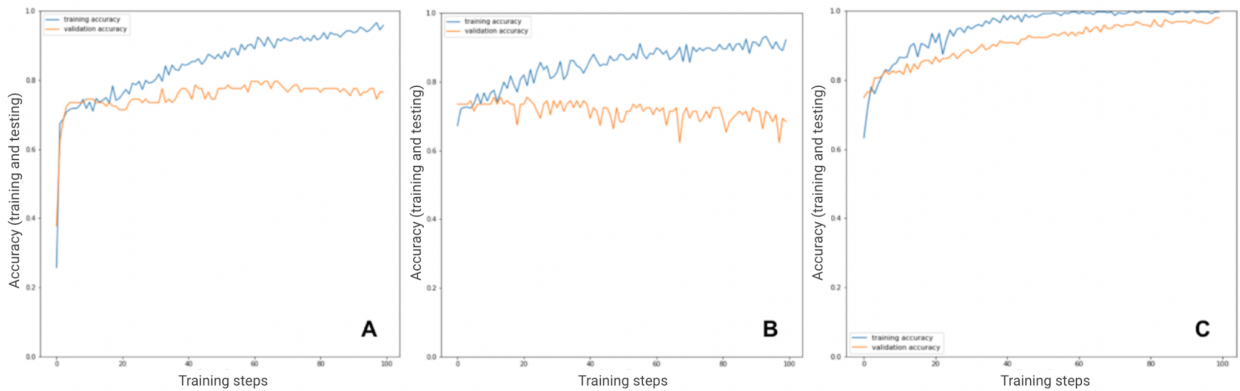

Figure 4: Training and testing accuracy curves for each of the models. (A) accuracy curves for symptoms based model; (B) accuracy curves for MRI images based model; (C) accuracy curves for hybrid model.

### 3.2 Results from MRI Scans Based Model

In this research, an MRI scans only based model was also developed, and its performance was measured with accuracy, precision, and recall. Initially, the model reached a training accuracy of 0.89, testing accuracy of 0.55, precision of 0.57, and recall of 0.57. In order to improve the initial model, image augmentation was applied by randomly rotating and cropping the scans (Shorten 2019). This data augmentation increased the size of our data set to a few thousand images, allowing for more training. After image augmentation, the model reached a training accuracy of 0.98, testing accuracy of 0.68, precision of 0.69, and recall of 0.74, as shown in Table 4. The difficulty the scans-based model had is reasonable since even with the current diagnosis procedure, MRI scans are not always useful. PD is very complex, and often, non-scans features like behavior analysis and surveys provide more information. Archer et al. 2019 achieved a testing accuracy of 0.90 with a support vector machine model. Archer's study had MRI scans from 1002 patients available to them, which is much more than the number of patients used in this research, 196. Furthermore, in this research, patients were classified into five categories instead of only two.

**3.3 Results from Hybrid Model**

In this research, a hybrid model that incorporates both the symptoms data and MRI scans was also developed. Initially, the hybrid model reached a training accuracy of 0.91, testing accuracy of 0.83, precision of 0.80, and recall of 0.81. To optimize the initial model, deep learning was applied. An extra layer was added before combining the outputs of the two composing elements. The layer was added in order to mix the individual symptoms-based and MRI scans-based results and to provide a more accurate final output. The deep learning model was more accurate, reaching a training accuracy of 0.99, testing accuracy of 0.98, precision of 0.98, and recall of 0.97, as shown in Table 4; there is no overfitting issue, since the testing accuracy is not much lower than the training accuracy. This hybrid architecture proves to be much more successful: its testing accuracy is 17% and 26% higher than the testing accuracies of the symptoms-based model and MRI scans-based model, respectively. The hybrid's success means that both assessment scores and MRI scans are useful in the diagnosis of PD. Since the hybrid model computes the weighted sums of the features, the two forms of information complement each other and help give a more accurate output. This is the first report that combines large-scale amounts of both forms of data to diagnose PD.

Additionally, the hybrid model can accurately diagnose Parkinson's disease even during its early stages. The model's stage-by-stage analytics were analyzed, and as displayed in Figure 5, the hybrid model achieved high precision, recall, and F1 scores across all five stages, including the early ones. For stage 1 detection, the model reached a 0.82 precision score, 0.94 recall score, and 0.88 F1 score; for stage 2 detection, the model reached a 0.98 precision score, 0.99 recall score, and 0.98 F1 score; for stage 3 detection, the model reached a 0.91 precision score, 0.83 recall score, and 0.87 F1 score; and for stage 4 detection, the model reached a 0.50 precision score, 0.50 recall score, and 0.50 F1 score. The metrics in stage 4 are lower since there were very few patients in this category and little data for the model to train on.

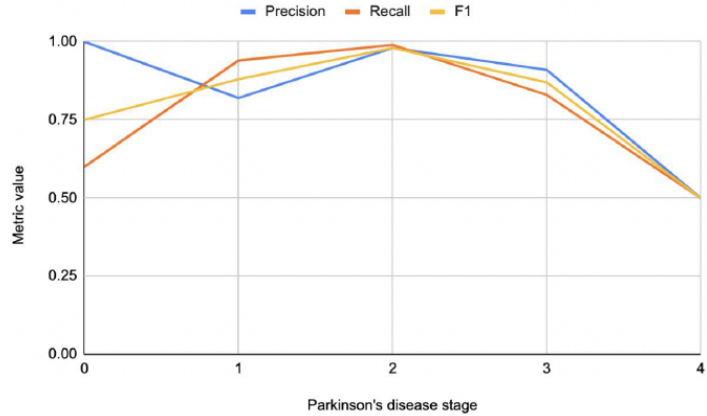

Figure 5: Graph of the precision, recall, and F1 scores metrics for the hybrid model across each of the 0-4 severity stages.

Parkinson's disease patients in early stages are usually harder to diagnose since motor symptoms have not yet developed and many tests that depend on physical developments are not reliable. However, early detection is critical, and if patients can begin treatment as soon as possible, they can hopefully avoid more severe symptoms in the future. The hybrid model performed very well across stages one and two, which represent patients in the early stages.

The hybrid model caught 15% of the errors made by the symptoms-based model and the MRI scans-based model. Table 5 shows a sample case, where the hybrid model corrected the base models' errors. The base models incorrectly classified by only a small margin, so with their complementary predictions,ced the hybrid model could give a more reliable prediction.

| Model | Predicted | Prediction Score [0, 1, 2, 3, 4] |
|---|---|---|
| Symptoms Based | 2 | [0.02, 0.45, 0.51, 0.00, 0.02] |
| MRI Scans Based | 3 | [0.02, 0.41, 0.12, 0.43, 0.02] |
| Hybrid | 1 | [0.10, 0.67, 0.19, 0.04, 0.01] |
| **Expected Category: 1** | | |

Table 5: Sample early stage case where the hybrid model corrected the base model's errors and gave more reliable predictions.

## 4. Conclusions

Based on the results, this research draws the following conclusions. First, we achieved our goal of designing a highly accurate deep learning algorithm with both test assessment scores and MRI scans. The model can successfully diagnose and classify Parkinson's disease patients into five categories. Second, assessments seem to be a better determiner of a patient's condition than MRI scans. The assessment score-based model was considerably more successful, which may mean MRI scans are simply too difficult to analyze or do not provide as much useful information for a diagnosis. Third, the novel procedure with the hybrid model has proven to be promising. The model was still successful in classifying patients into five categories. It is more accurate than both the assessment score-based model and the MRI scans-based model and analyzes the two forms of data efficiently. Finally, the hybrid model has strong early detection of Parkinson's disease, as it performs well for patients in stages 1 and 2. This is especially beneficial for patients and doctors since it is critical to diagnose the disease and to begin treatment as early as possible (Rees 2018).

Compared to previous studies, our hybrid model is very strong. In addition, because we utilized so many symptom assessments, our symptoms-based model performed better than others, such as that of Archer et al. 2019. Our MRI scans-based model was likely weak because we had a limited amount of data; for example, Archer et al. 2019 achieved a testing accuracy of 0.90 with a support vector machine model, but they used scans from 1002 patients, which is much more than the number of patients used in this research, 196. In addition, in this research, patients were classified into five categories instead of only two.

There are advantages and disadvantages to each model. Symptom assessments are easy to complete, inexpensive, and accessible to most patients, and the symptoms-based model still reaches a reliably high accuracy of 0.77; however, it is the worst-performing model in recall and precision, which means it has a high number of dangerous false positives and false negatives, and is significantly less accurate than the hybrid technology. The MRI scans-based model has a higher recall, 0.74, than the symptoms-based model, and can likely be improved greatly with more data; however, MRI scans are expensive and time-consuming, and many underdeveloped areas do not have scanning technology (Kavosi 2021). The hybrid model exceeds significantly in all metrics, with an accuracy of 0.94 and high recall and precision, and metrics are consistent across all stages; however, it requires both forms of data.

In the future, we can work to achieve higher accuracies and to provide even more to PD doctors and patients. First, the logistic regression model can be greatly optimized if we use more assessments, since each test measures different symptoms that can be present in Parkinson's disease patients. Second, we can similarly optimize the hybrid model with more data. Currently, the number of MRI scans is more limiting. Third, we can implement models that produce continuous outputs as diagnoses instead of just classifying them into categories. Finally, this hybrid novel procedure likely has not been extended to other areas of the medical field, and it can contribute greatly to the diagnoses of other diseases. With a few adaptations, the model can be a strong deep learning tool for brain conditions like Alzheimer's disease or atypical parkinsonism.

**Acknowledgments**

I would like to thank Dr. Eric Nelson at The Harker School for sponsoring my project and for providing me with the guidance necessary to conduct this research. I am incredibly grateful for the opportunity and all that I have learned throughout the process.

**References**

Abel, S., Kolind, S. (2019). *Improving parkinsonism diagnosis with machine learning.* https://www.thelancet.com/journals/landig/article/PIIS2589-7500(19)30107-4/fulltext.

Akoglu, H. (2018). *User's guide to correlation coefficients*. https://www.ncbi.nlm.nih.gov/pmc/articles/PMC6107969/.

American Association of Neurological Surgeons. (2021). *Parkinson's Disease*.

Archer, D., Bricker, J., Chu, W., et al. (2019). *Development and Validation of the Automated Imaging Differentiation in Parkinsonism (AID-P): A Multi-Site Machine Learning Study*. https://pubmed.ncbi.nlm.nih.gov/32259098/.

BrainSuite. http://brainsuite.org/.

Brant-Zawadzki, M., Gillan, G. D., Nitz, W. R. (1992). *MP RAGE: a three-dimensional, T1-weighted, gradient-echo sequence--initial experience in the brain*. https://pubmed.ncbi.nlm.nih.gov/1535892/.

Chakraborty, S., Aich, S., and Kim, H. (2020). *Detection of Parkinson's Disease from 3T T1 Weighted MRI Scans Using 3D Convolutional Neural Network.* https://www.ncbi.nlm.nih.gov/pmc/articles/PMC7345307/.

Christodoulou, E., Ma, J., Collins, G., et al. (2019). *A systematic review shows no performance benefit of machine learning over logistic regression for clinical prediction models*. https://www.sciencedirect.com/science/article/abs/pii/S0895435618310813.

*Diagnosing Parkinson's*. American Parkinson Disease Association. https://www.apdaparkinson.org/what-is-parkinsons/diagnosing/.

Doty, R., Shaman, P., Kimmelman, C., et al. (1984). *University of Pennsylvania Smell Identification Test: a rapid quantitative olfactory function test for the clinic.* https://pubmed.ncbi.nlm.nih.gov/6694486/.

Esmaeilzadeh, S., Yang, Y., Adeli, E. (2018). *End-to-End Parkinson Disease Diagnosis using Brain MR-Images by 3D-CNN*. https://arxiv.org/abs/1806.05233.

Goetz, C., Fahn, S., Martinez-Martin, P., et al. (2008). *The MDS-sponsored Revision of the Unified Parkinson's Disease Rating Scale.* International Parkinson and Movement Disorder Society. https://www.movementdisorders.org/MDS-Files1/PDFs/Rating-Scales/MDS-UPDRS_English_FINAL_Updated_August2019.pdf.

Goetz, C., Poewe, W., Rascol, O., et al. (2004). *Movement Disorder Society Task Force report on the Hoehn and Yahr staging scale: status and recommendations*. https://pubmed.ncbi.nlm.nih.gov/15372591/.

Haller, S., Badoud, S., Nguyen, D., et al. (2012). *Individual Detection of Patients with Parkinson Disease using Support Vector Machine Analysis of Diffusion Tensor Imaging Data: Initial Results.* http://www.ajnr.org/content/33/11/2123.

Heim, B., Krismer, F., De Marzi, R., et al. (2017). *Magnetic resonance imaging for the diagnosis of Parkinson's disease.* https://www.ncbi.nlm.nih.gov/pmc/articles/PMC5514207/.

Hoehn, M., Yahr, M. (1967). *Parkinsonism: onset, progression, and mortality.* https://n.neurology.org/content/neurology/17/5/427.full.pdf.

Kavosi, Z., Sadeghi, A., Lotfi, G. (2021). *The inappropriateness of brain MRI prescriptions: a study from Iran*. https://resource-allocation.biomedcentral.com/articles/10.1186/s12962-021-00268-6.

Marras, C., Beck, J. C., Bower, J.H. (2018). *Prevalence of Parkinson's disease across North America*. https://www.ncbi.nlm.nih.gov/pmc/articles/PMC6039505/.

Mozhdehfarahbakhsh, A., Chitsazian, S., Chakrabarti, P., et al. (2021). *An MRI-based Deep Learning Model to Predict Parkinson's Disease Stages.* https://www.medrxiv.org/content/10.1101/2021.02.19.21252081v1.full.

Nasreddine, Z., Phillips, N., Bédirian, V., et al (2005). *The Montreal Cognitive Assessment, MoCA: a brief screening tool for mild cognitive impairment*. https://pubmed.ncbi.nlm.nih.gov/15817019/.

National Institute of Health's National Institute of Aging. (2017). *Parkinson's Disease*. https://www.nia.nih.gov/health/parkinsons-disease.

O'Shea, K., Nash, R. (2015) *An Introduction to Convolutional Neural Networks*. https://www.researchgate.net/publication/285164623_An_Introduction_to_Convolutional_Neural_Networks.

Pagano, G., Niccolini, F., Politis, M. (2016). *Imaging in Parkinson's disease*. https://www.ncbi.nlm.nih.gov/pmc/articles/PMC6280219/.

Parkinson's Progression Markers Initiative. http://www.ppmi-info.org/.

Postuma, R. B., Berg, D., Stern, M. (2015). *MDS clinical diagnostic criteria for Parkinson's disease*. https://pubmed.ncbi.nlm.nih.gov/26474316/.

Prashanth, R., Roy, S., Mandal, P., et al. (2016). *High-Accuracy Detection of Early Parkinson's Disease through Multimodal Features and Machine Learning*. https://pubmed.ncbi.nlm.nih.gov/27103193/.

Rees, R., Acharya, A., Schrag, A., et al. (2018). *An early diagnosis is not the same as a timely diagnosis of Parkinson's disease*. https://www.ncbi.nlm.nih.gov/pmc/articles/PMC6053699/.

Rizzo, G., Copetti, M., Arcuti, S., et al. (2016). *Accuracy of clinical diagnosis of Parkinson disease: A systematic review and meta-analysis*. https://pubmed.ncbi.nlm.nih.gov/26764028/.

Saha, R. (2019) *Classification of Parkinson's Disease Using MRI Data and Deep Learning Convolution Neural Networks*. https://lib.dr.iastate.edu/cgi/viewcontent.cgi?article=1305&context=creativecomponents.

Sandler, M., Howard, A. (2018). *MobileNetV2: The Next Generation of On-Device Computer Vision Networks*. Accessed 26 February 2021. https://ai.googleblog.com/2018/04/mobilenetv2-next-generation-of-on.html.

Shorten, C., Khoshgoftaar, T. (2019). *A survey on Image Data Augmentation for Deep Learning*. https://journalofbigdata.springeropen.com/articles/10.1186/s40537-019-0197-0.

Sivaranjini, S., Sujatha, C. M. (2019). *Deep learning based diagnosis of Parkinson's disease using convolutional neural network.*
https://link.springer.com/article/10.1007/s11042-019-7469-8.

Stiasny-Kolster, K., Mayer, G., Schafer, S., et al. (2007). *The REM sleep behavior disorder screening questionnaire--a new diagnostic instrument*.
https://pubmed.ncbi.nlm.nih.gov/17894337/.

Visser, M., Marinus, J., Stiggelbout, A., et al. (2004). *Assessment of Autonomic Dysfunction in Parkinson's Disease: The SCOPA-AUT.*
https://www.movementdisorders.org/MDS-Files1/Education/PDFs/Visser_et_al-2004-Movement_Disorders.pdf.

Washburn, R., Smith, K., Jette, A., et al. (1993). *The Physical Activity Scale for the Elderly (PASE): development and evaluation*. https://pubmed.ncbi.nlm.nih.gov/8437031/.

Yang, W., Hamilton, J., Kobil, C., et al. (2020). *Current and projected future economic burden of Parkinson's disease in the U.S.* https://www.nature.com/articles/s41531-020-0117-1.
16